\documentclass[preprint2,times]{aastex62}
 \hypersetup{
     urlcolor=blue,
 }
\usepackage{multirow}
\usepackage{textcomp}
\usepackage{amsmath}
\usepackage{gensymb}

\shorttitle{Unambiguous Evidence of Coronal Implosions}
\shortauthors{Wang et al.}

\begin{document}

\title{Unambiguous Evidence of Coronal Implosions During Solar Eruptions and Flares}

\correspondingauthor{Juntao Wang}
\email{j.wang.4@research.gla.ac.uk}

\author[0000-0001-9268-2966]{Juntao Wang}
\affil{SUPA, School of Physics and Astronomy, University of Glasgow, Glasgow G12 8QQ, UK}

\author[0000-0002-4819-1884]{P. J. A. Sim\~{o}es}
\affil{SUPA, School of Physics and Astronomy, University of Glasgow, Glasgow G12 8QQ, UK}

\author[0000-0001-9315-7899]{L. Fletcher}
\affil{SUPA, School of Physics and Astronomy, University of Glasgow, Glasgow G12 8QQ, UK}

\begin{abstract}
In the implosion conjecture, coronal loops contract as the result of magnetic energy release in solar eruptions and flares. However, after almost two decades, observations of this phenomenon are still rare, and most of previous reports are plagued by projection effects so that loop contraction could be either true implosion or just a change in loop inclination. In this paper, to demonstrate the reality of loop contractions in the global coronal dynamics, we present four events with the continuously contracting loops in an almost edge-on geometry from the perspective of SDO/AIA, which are free from the ambiguity caused by the projection effects, also supplemented by contemporary observations from STEREO for examination. In the wider context of observations, simulations and theories, we argue that the implosion conjecture is valid in interpreting these events. Furthermore, distinct properties of the events allow us to identify two physical categories of implosion. One type demonstrates a rapid contraction at the beginning of the flare impulsive phase, as magnetic free energy is removed rapidly by a filament eruption. The other type, which has no visible eruption, shows a continuous loop shrinkage during the entire flare impulsive phase which we suggest manifests the ongoing conversion of magnetic free energy in a coronal volume. Corresponding scenarios are described, which can provide reasonable explanations for the observations. We also point out that implosions may be suppressed in cases when a heavily-mass-loaded filament is involved, possibly serving as an alternative account for their observational rarity.
\end{abstract}


\section{Introduction} \label{intro}
Solar eruptions and flares are two main manifestations of magnetic energy release in the corona of the Sun. \cite{hud2000} conjectured that a new phenomenon termed ``implosion'' would accompany these energy release processes, based on the assumption of the dominance of Lorentz force in the coronal dynamics, and the equivalence of magnetic energy and magnetic pressure. The conjecture reads ``During a transient, the coronal field lines must contract in such a way as to reduce $\int_V(B^2/8\pi)dV$''. Though it was proposed almost two decades ago, observations of such field implosion phenomena are still rare, compared to numerous eruptions and flares observed.

Remarkable coronal loop contractions in extreme ultraviolet at the periphery of active regions, with speeds of tens to hundreds of km/s, were reported in a few events ranging from Geostationary Operational Environmental Satellite (GOES) class B to X \citep{liu2009,liu2010,gos2012,liu2012,sun2012,sim2013,yan2013,kus2015,wan2016}. As these peripheral loop contractions were always observed face-on and accompanied by eruptions from central magnetic structures (like a filament or an arcade eruption), the possibility could not be ruled out that apparent contraction is a projection effect due to inclination of the loop plane pushed by the erupting structure, rather than a real contraction (from our survey experience, loop inclining is indeed more commonly observed when the loops are viewed with an edge-on state at the solar limb, and even some of them do not restore back to their original locations). As far as we know, only \cite{pet2016} reported edge-on loop contractions in two active regions from the perspective of Solar TErrestrial RElations Observatory (STEREO) in 195 {\AA}, but due to the short interval of the process and the long cadence ($\sim5$ min), the dynamics was not persistently revealed and not clear enough to be well studied. In addition, both of the \cite{pet2016} events show dramatic eruptions, but in this paper we also show a new type of loop contractions observed edge-on without violent eruptions. The argument that the contracting loops do not restore to their original positions after the eruptions \citep{liu2012,gos2012,sim2013,wan2016}, and evidence from NLFFF extrapolations \citep{wan2016} has been used to try to substantiate the reality of the contracting motion, but the doubt that it could be a projection effect can still not be completely excluded, and the ambiguity remains.

In some of the events above, dramatic oscillations were noticed during or after the loop contractions \citep{liu2010,gos2012,liu2012,sun2012,sim2013}. \cite{rus2015} considered a one loop system as a harmonic oscillator, showing that the contracting and oscillating behaviours can be reproduced by the change in loop equilibrium position due to magnetic energy release underneath, in agreement with the implosion conjecture. \cite{pas2017} included a displacement term for the changing equilibrium position from \cite{rus2015} for coronal seismology analysis, and only the fundamental kink mode exists associated with the loop contraction in \cite{sim2013}. \cite{liu2010} suggested that the interaction between the contracting loops and surrounding ones may also make them oscillate. The model of an isolated simple harmonic oscillator cannot properly describe the dynamics of a continuum medium, where many magnetic strands will interact with each other if not in phase, so a full magnetohydrodynamic (MHD) treatment may be needed for a more accurate description of the dynamics.

A longitudinal field decrease or horizontal field enhancement near the polarity inversion line in the photospheric magnetograms has been detected during many events, especially eruptive flares \citep{sud2005,pet2010,wan2010,gos2012,pet2012,sun2012,sun2017}. The phenomenon is often explained by the authors exploiting the implosion conjecture, because this predicts a more horizontal field as loops contract, which could probably propagate from the restructuring corona down to the photosphere during the impusive phase \citep{hud2008,fis2012}. However, the non-eruptive X3.1 flare in the famous active region 12192 did not show significant changes in its photospheric horizontal field \citep{sun2015,jiang2016}.

In several MHD simulations with a configuration in which a flux rope is anchored below a magnetic arcade, when the flux rope erupts outward, it can be seen that some of the peripheral unopened arcade field finally contracts\footnote{Depending on the location of the arcade field, the field would expand, incline and contract, or incline and contract, or directly contract.} toward the central erupting structure, leading to a shorter length compared to its pre-eruption state \citep{aul2005,gib2006,fan2007,rac2009,zuc2017,dud2017}. However, \cite{zuc2017} and \cite{dud2017} proposed an alternative explanation of the loop contraction in their simulation, using the analogy of vortices in the hydrodynamic situation (further discussed in Section~\ref{model}). \cite{sar2017} recently carried out the first simulation focused on implosions, and found that oscillations of both kink and sausage modes can exist when the loops contract, and that loops in different plasma $\beta$ regimes may exhibit different dynamic behaviours. 

In this paper, we will present direct evidence of continuous implosion phenomena, with the observations shown in Section~\ref{obser}. Based on the main observational properties, Section~\ref{discuss} will demonstrate the validity of the implosion conjecture, and categorize the observed implosions into two types, with corresponding models proposed. Conclusions are summarized in Section~\ref{conclusion}.

\section{Observations and Analyses} \label{obser}
We select four events, SOL2011-09-14T16:26 (C4.2), SOL2014-02-17T23:15 (C1.9), SOL2016-04-08T01:56 (B8.3), and SOL2016-11-22T23:45 (B6.0), for analysis, which are located in active regions NOAA 11290 (S17W47), 11978 (N05W89), 12529 (N09E88), and 12612 (N11E89). Hereafter, for convenience, the four events are labelled as Event I, II, III, and IV, respectively. They are all observed by both Solar Dynamics Observatory/Atmospheric Imaging Assembly (SDO/AIA) and STEREO A. The contracting arcades in these four events all have an almost edge-on geometry from the perspective of AIA, so the contributions to the loop dynamics from contraction and inclination can be clearly disentangled. The contracting loops observed by STEREO A in 195 {\AA} are very likely the same as that viewed from AIA in 193 {\AA} (for Event IV the contracting structures in 171 {\AA} are similar to that in 193 {\AA}), because (1) these two wave bands share similar observing temperature $\sim1.5\times10^6$ K; (2) the structures observed from the two perspectives show expected positions and geometry according to the relative positions of SDO and STEREO; and (3) there are good temporal correspondences between the contracting motions captured by the two observatories. AIA images and photospheric magnetograms from Helioseismic and Magnetic Imager (HMI) for Event I have been processed by the standard software \citep{boe2012}, and supplementary images from STEREO A via secchi\_prep.pro \citep{how2008}.

\subsection{Event I: SOL2011-09-14T16:26}

\begin{figure*}
\hspace{0cm}\includegraphics[scale=1]{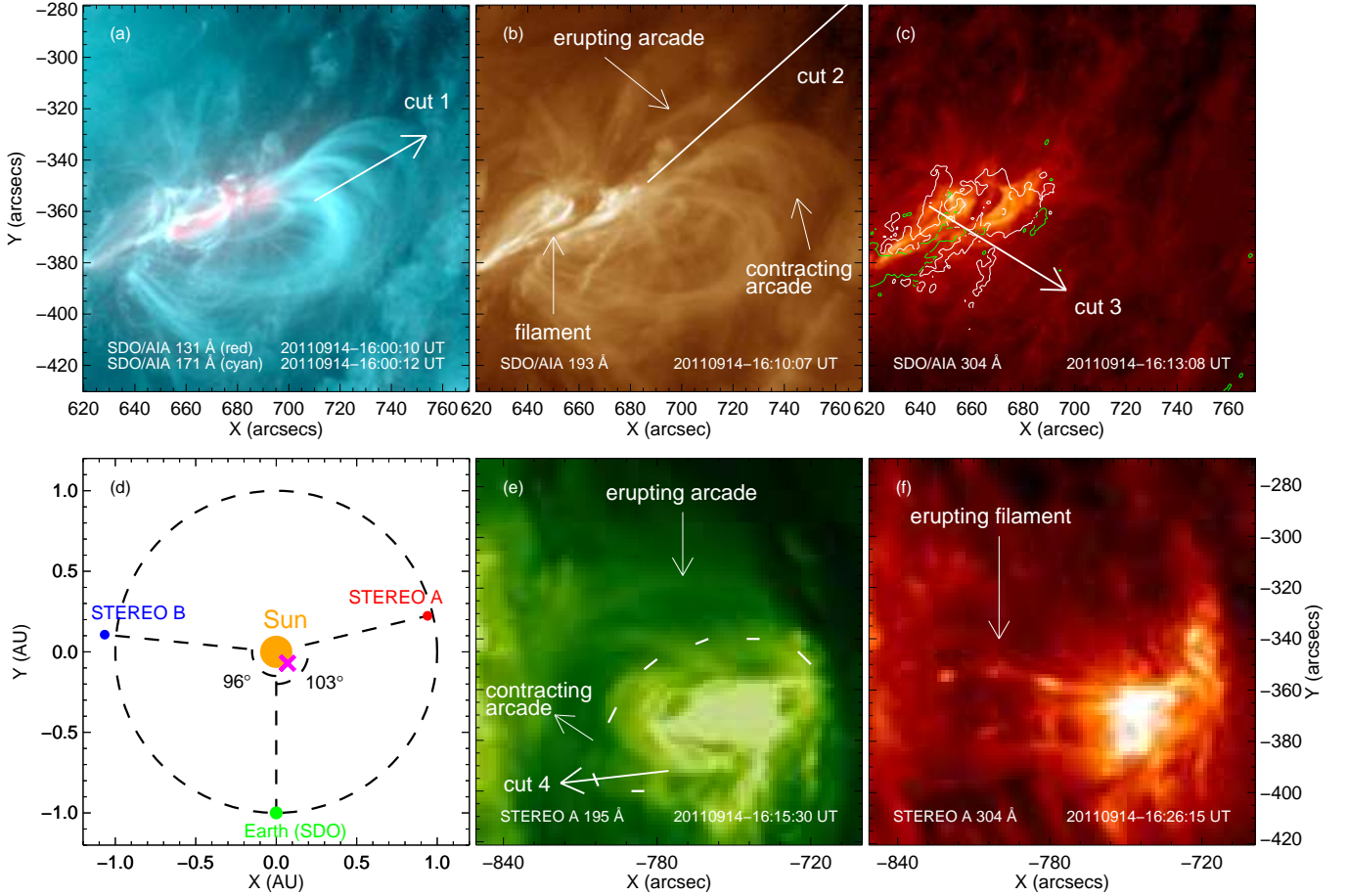}
\caption{\label{cut1}Images for Event I: SOL2011-09-14T16:26. (a)-(c) observed from the perspective of AIA. 131 {\AA} is red, and 171 {\AA} cyan in (a) (hereafter for composite images, cyan always represents a low temperature band, like 171 or 193 {\AA}, and the hot 131 {\AA} is always set to red). (d) relative positions of SDO and STEREO. The magenta cross shows the longitudinal position of the event. (e)-(f) observed from the perspective of STEREO A. The dashed line in (e) illustrates the location and shape of the contracting arcade. Cuts 1-4 are used for the timeslices in Figure~\ref{ts1}. The arrowhead of cut 2 is beyond the image edge. An animation of this figure is available at \url{http://researchdata.gla.ac.uk/598/}. An additional animation Event\_I.mov showing the correspondence between the structures viewed from AIA and STEREO A is also provided at the link, which exploits JHelioviewer \citep{mul2017}.}
\end{figure*}

\begin{figure*}
\hspace{0cm}\includegraphics[scale=1]{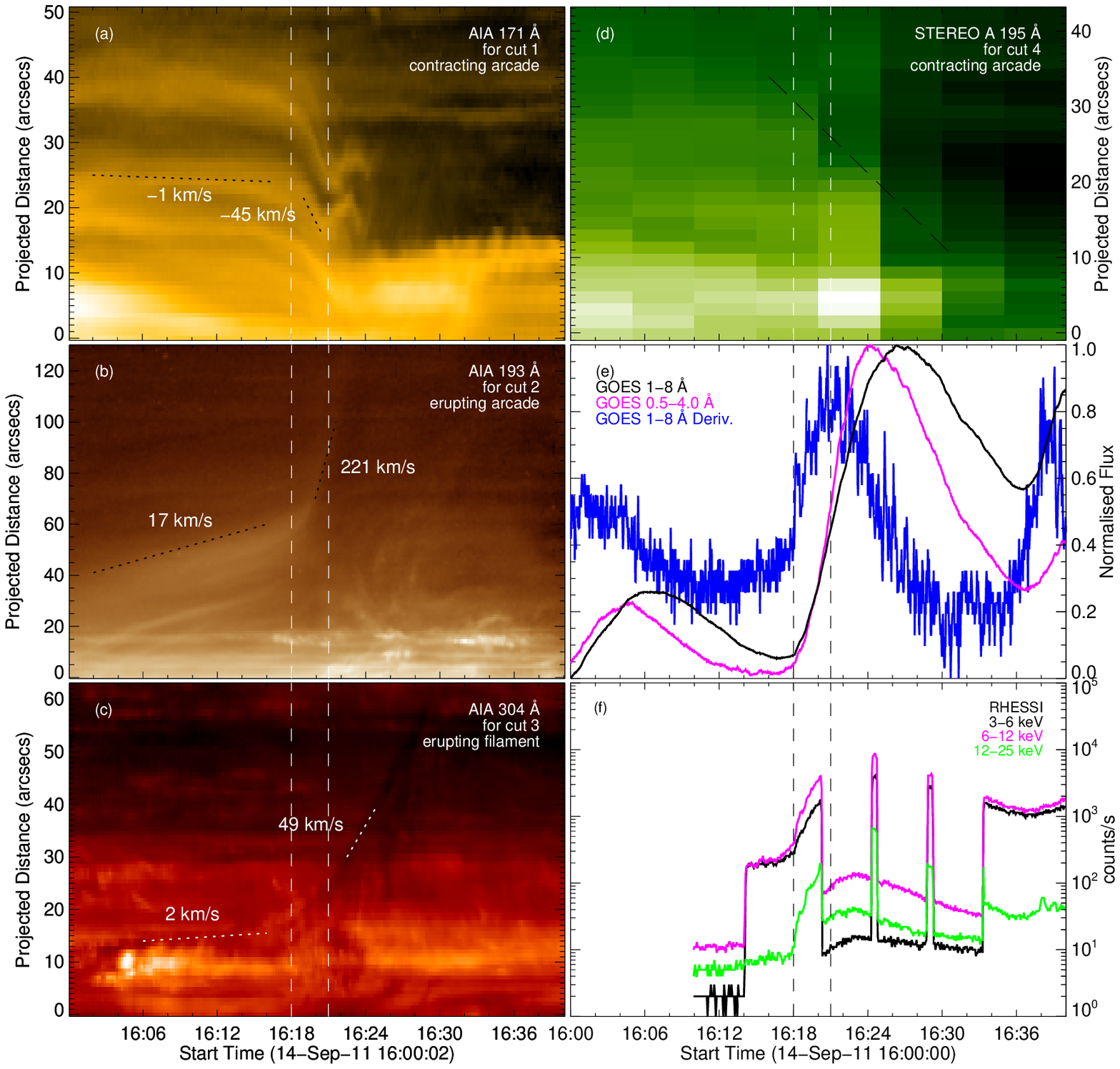}
\caption{\label{ts1}(a)-(d) Timeslices for dynamic features in event I. The sampling time of STEREO A 195 {\AA} in (d) starts from the beginning of each timeslice, with an exposure duration $\sim8$ s, and the long-dashed line shows the rough contraction trend but means an uncertain contraction speed because of the long sampling cadence $\sim5$ min and few sampling points. (e)-(f) GOES and RHESSI light curves, respectively. The two vertical dashed lines across the figure shows the time interval of the arcade contraction.}
\end{figure*}

Event I is shown in Figure~\ref{cut1} and the accompanying animation, with both AIA and STEREO A observations. AIA observes the contracting arcade (hereafter we call it arcade I) from the side with a nearly horizontal geometry (Figure~\ref{cut1}(a) and (b)), while STEREO A looks at it from the top with the loop plane having $\sim45$\degree~ with respect to the line of sight (Figure~\ref{cut1}(e)). A filament is located low in the corona (Figure~\ref{cut1}(c)). As it is destabilised and erupts outward (Figure~\ref{cut1}(f)), another arcade structure  (hereafter arcade II) passes from beneath arcade I and erupts (Figure~\ref{cut1}(b)). Meanwhile, arcade I contracts towards the space left by the erupting filament and arcade II. The motion of contraction is unambiguous, which is evidenced by the accompanied animation. Oscillation follows and finally most of the loops of arcade I disappear.

Figure~\ref{ts1}(a)-(d) show the timeslices created along the cuts 1-4 chosen in Figure~\ref{cut1}, respectively, presenting the detailed dynamics of the corresponding features along the cuts. The major contraction of arcade I (in the interval between the two dashed lines) starts as the filament and arcade II erupt, though they already have similar but weaker behaviours before this time interval. This major contraction interval also corresponds to the rise of the impulsive phase, which is illustrated by the GOES 1-8 {\AA} derivative in Figure~\ref{ts1}(e) and the light curve of RHESSI 12-25 keV in Figure~\ref{ts1}(f). After the major contraction, the loops of arcade I oscillate and most of them disappear (Figure~\ref{ts1}(a)), though the filament and arcade II still continue to move outward rapidly(Figure~\ref{ts1}(b) and (c)). We note that the contraction speed of arcade I is always much smaller than the eruption speeds of arcade II and also the filament. The filament eruption speed is underestimated in Figure~\ref{ts1} because of projection, and can be more accurately estimated to be $\sim150~\rm km~s^{-1}$, by considering the time interval between 16:18:00 UT (the start time of the filament eruption from Figure~\ref{ts1}(c)) and 16:26:15 UT (Figure~\ref{cut1}(f)), and the travel distance $\sim100$ arcsecs in Figure~\ref{cut1}(f). The final contraction distance of arcade I is also much smaller than the final eruption distances of the filament and arcade II.

\subsection{Event II: SOL2014-02-17T23:15}

\begin{figure*}
\hspace{0cm}\includegraphics[scale=1]{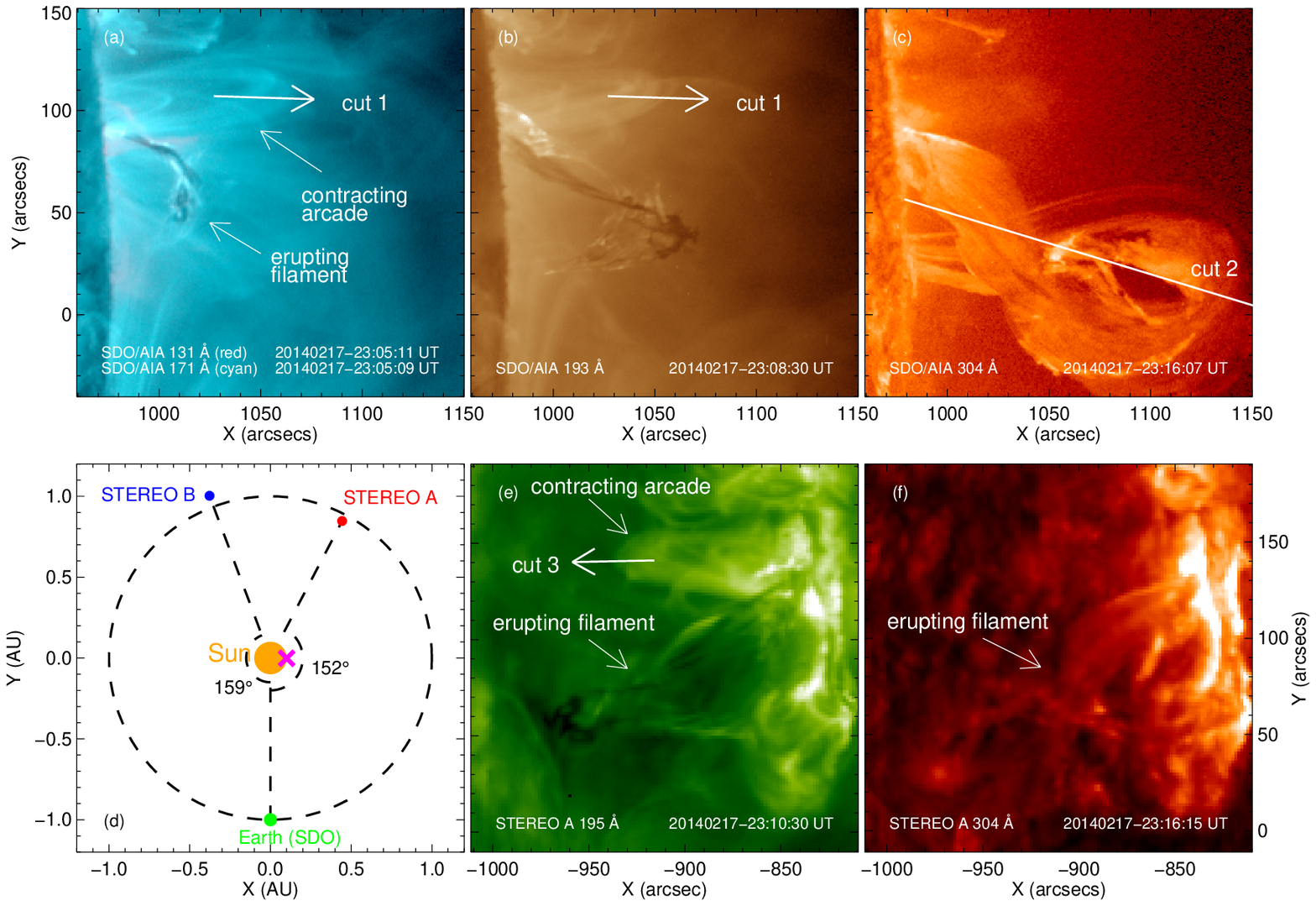}
\caption{\label{cut2}Images for Event II SOL2014-02-17T23:15. (a)-(c) observed from the perspective of AIA. 131 {\AA} is red, and 171 {\AA} cyan in (a). (d) relative positions of SDO and STEREO. The magenta cross shows the longitudinal position of the event. (e)-(f) observed from the perspective of STEREO A. Cuts 1-3 are used for the timeslices in Figure~\ref{ts2}. The arrowhead of cut 2 is beyond the image edge. An animation of this figure is available at \url{http://researchdata.gla.ac.uk/598/}. An additional animation Event\_II.mov showing the correspondence between the structures viewed from AIA and STEREO A is also provided at the link, which exploits JHelioviewer \citep{mul2017}.}
\end{figure*}

\begin{figure}
\hspace{0cm}\includegraphics[scale=1.15]{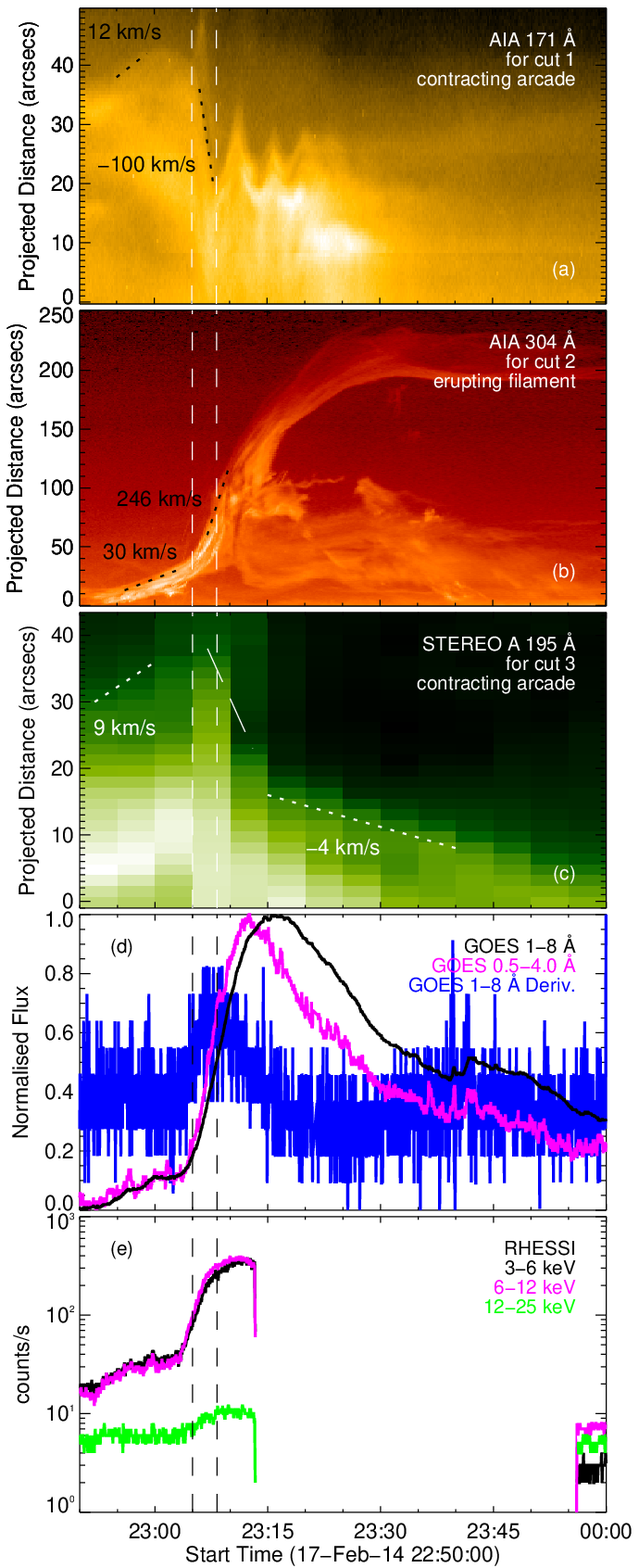}
\caption{\label{ts2}(a)-(c) Timeslices for dynamic features in event II. The sampling time of STEREO A 195 {\AA} in (C) starts from the beginning of each timeslice, with an exposure duration $\sim8$ s, and the long-dashed line shows the rough contraction trend but means an uncertain contraction speed because of the long sampling cadence $\sim5$ min and few sampling points. (d)-(e) GOES and RHESSI light curves, respectively. The two vertical dashed lines across the figure shows the time interval of the arcade contraction.}
\end{figure}

Figures~\ref{cut2} and \ref{ts2} are constructed similarly to Figures~\ref{cut1} and \ref{ts1}, respectively. Event II is located on the limb with a more favourable perspective, making the contraction of the arcade clearer. Seen from the accompanying animation, first the filament lies close to the solar surface, with the arcade overlying its northern end. Then they expand upward simultaneously up to around 23:05 UT (Figure~\ref{cut2}(a)). As the filament starts to writhe along with its southwestward eruption (Figure~\ref{cut2}(b)), the arcade begins to contract and the northern end of the filament seems to be pushed downward to the solar surface. In the end the arcade oscillates and gradually disappears. 

Similar to Event I, the major arcade contraction coincides with the beginning of the filament eruption and the rise stage of the impulsive phase, and the arcade contracts more slowly and over a much smaller distance than the filament erupts (Figure~\ref{ts2}). Event II differs from Event I in that before the major contraction, the arcade in Event II shows slow expansion rather than slow contraction as in Event I.

\subsection{Event III: SOL2016-04-08T01:56}

\begin{figure*}
\hspace{0cm}\includegraphics[scale=1]{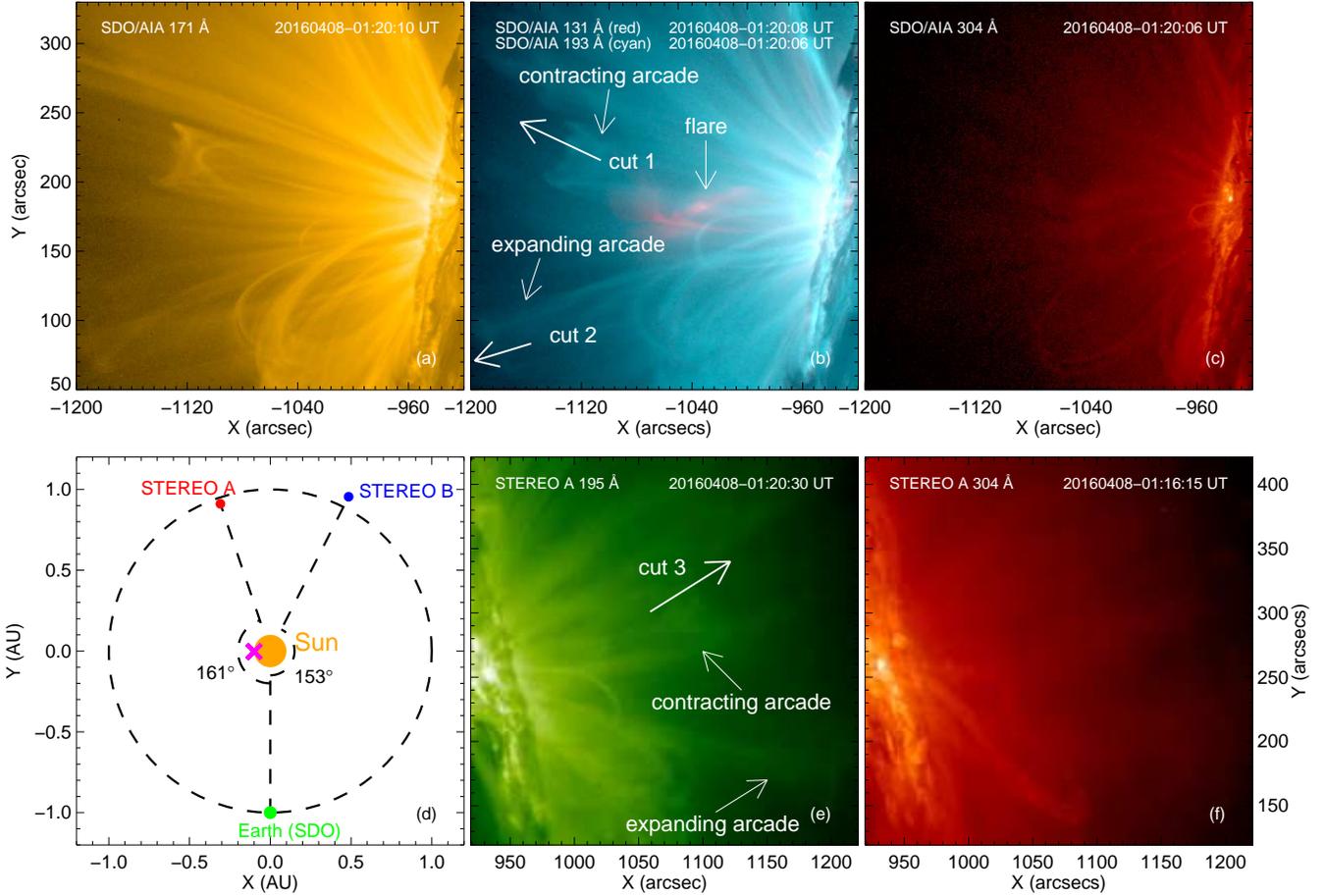}
\caption{\label{cut3}Images for Event III SOL2016-04-08T01:56 B8.3. (a)-(c) observed from the perspective of AIA. 131 {\AA} is red, and 193 {\AA} cyan in (b). (d) relative positions of SDO and STEREO. The magenta cross shows the longitudinal position of the event. (e)-(f) observed from the perspective of STEREO A. Cuts 1-3 are used for the timeslices in Figure~\ref{ts3}. An animation of this figure is available at \url{http://researchdata.gla.ac.uk/598/}.}
\end{figure*}

\begin{figure}
\hspace{0cm}\includegraphics[scale=1.15]{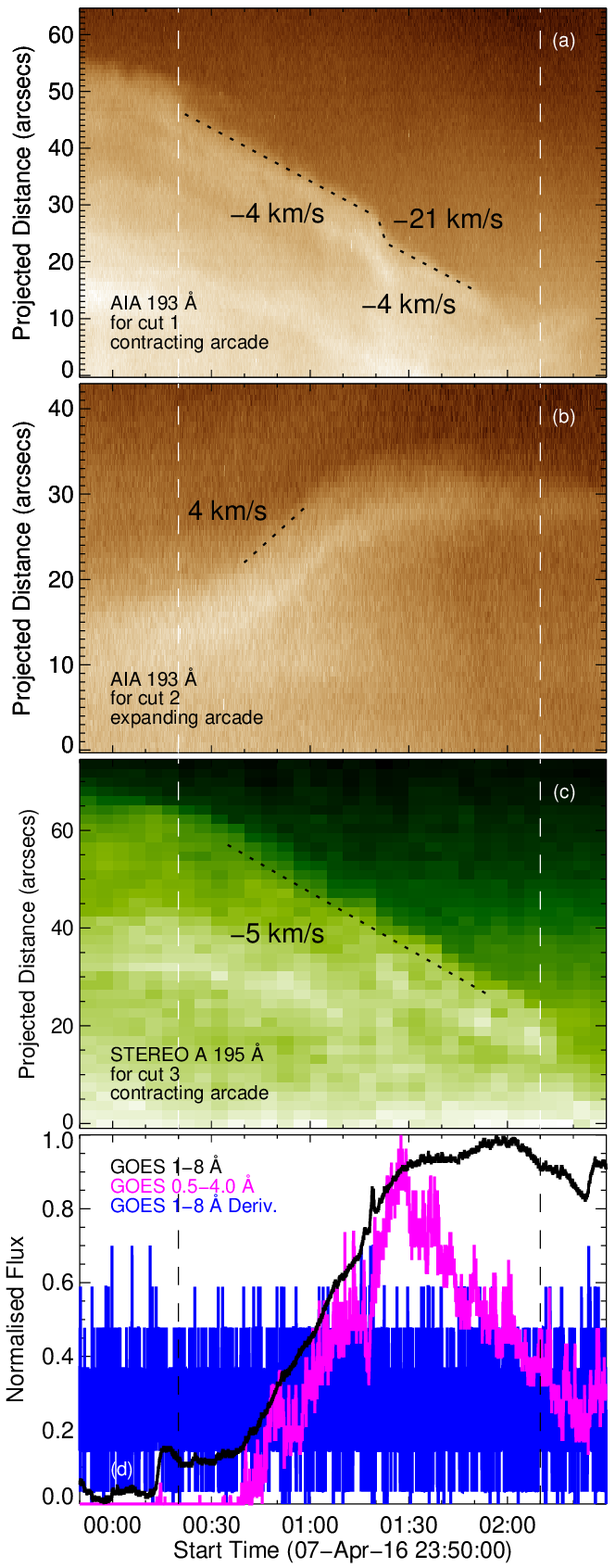}
\caption{\label{ts3}(a)-(c) Timeslices for dynamic features in event III. The sampling time of STEREO A 195 {\AA} in (C) starts from the beginning of each timeslice, with an exposure duration $\sim8$ s. (d) GOES light curves. The two vertical dashed lines across the figure shows the time interval of the arcade contraction.}
\end{figure}

AIA and STEREO A observe the contracting arcade in Event III from opposite sides (Figure~\ref{cut3}(a), (b) and (e)). The arcade contracts as a flare underneath happens (Figure~\ref{cut3}(b)). Strangely, neither AIA nor STEREO observations, which together have a wide temperature coverage (including cool 304 {\AA}, warm 171 , 193 and 195 {\AA}, and hot 131 {\AA}) show any signature of violent arcade or filament eruptions as seen in Event I and II. There is only another arcade in the south expanding outward to a small extent (Figure~\ref{cut3}(b)). The arcade in the north fades into the flaring region at the end with no obvious oscillation detected.

Figure~\ref{ts3}(a) shows that the speed of the long-duration arcade contraction is only a few $\rm km~s^{-1}$, which is slow but real, rather than caused by solar rotation, because there are surrounding static loops as a reference (see the accompanying animation). And interestingly, an abrupt acceleration in the contraction occurs at around 01:20 UT, which coincides with a sudden increase or a spike in GOES 1-8 {\AA} light curve (Figure~\ref{ts3}(d)). It seems that the contraction of the arcade is quite sensitive to the flare. Though the Neupert effect is not notable here, the contraction process has already continued past the peak of the GOES 1-8 {\AA} flux, which means that the arcade contraction spans the entire impulsive phase. This is unlike the situations in Events I and II where the contraction is localized in time to the rise of the impulsive phase. The expansion speed of the arcade in the south is also very small (Figure~\ref{ts3}(b)), comparable to the contraction speed of the arcade in the north, but it only persists for about half of the contraction interval, which results in an expansion distance of around half of the contraction distance.   

\subsection{Event IV: SOL2016-11-22T23:45}

\begin{figure*}
\hspace{0cm}\includegraphics[scale=1]{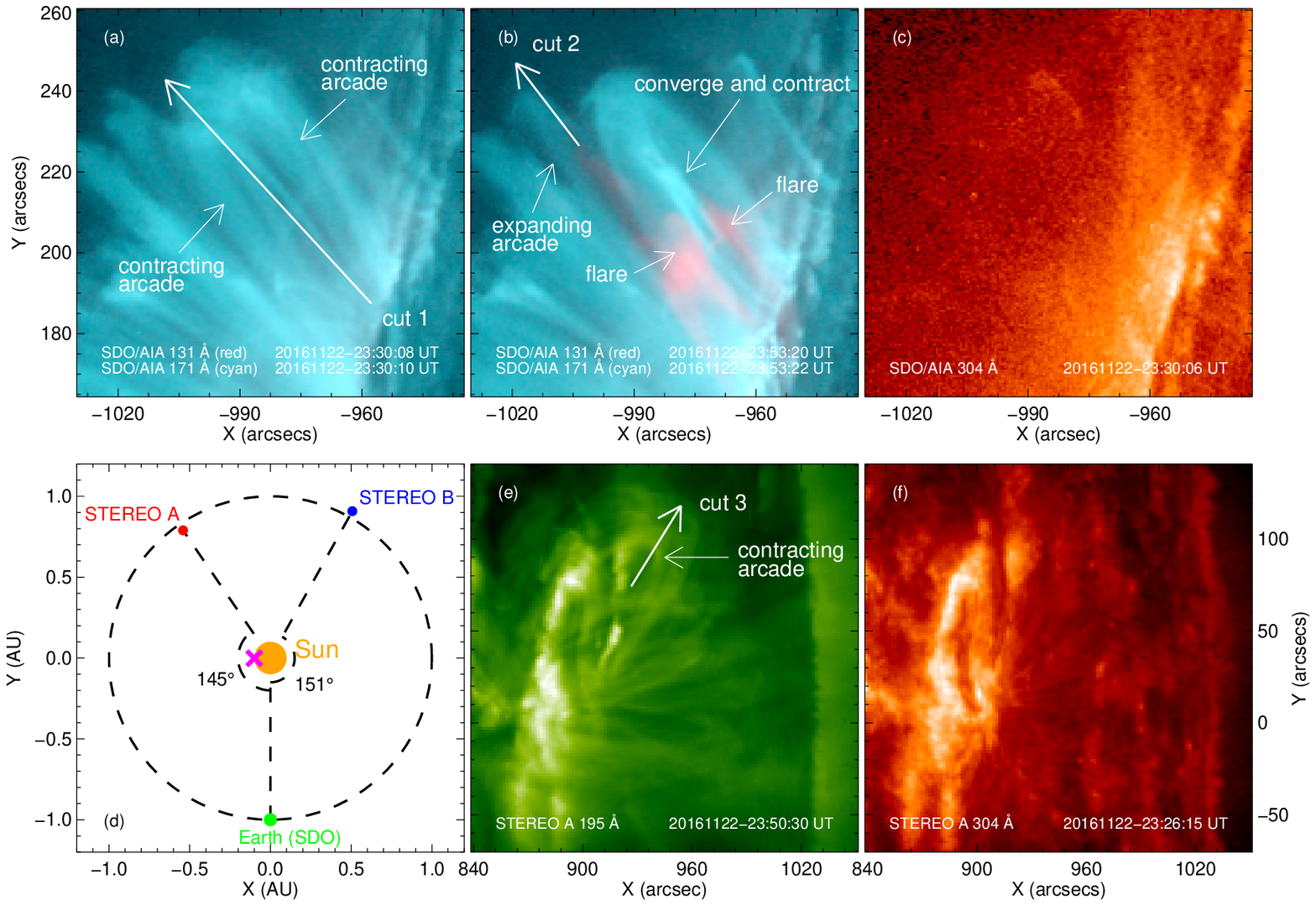}
\caption{\label{cut4}Images for Event IV SOL2016-11-22T23:45 B6.0. (a)-(c) observed from the perspective of AIA. 131 {\AA} is red, and 171 {\AA} cyan in (a) and (b). (d) relative positions of SDO and STEREO. The magenta cross shows the longitudinal position of the event. (e)-(f) observed from the perspective of STEREO A. Cuts 1-3 are used for the timeslices in Figure~\ref{ts4}. An animation of this figure is available at \url{http://researchdata.gla.ac.uk/598/}. An additional animation Event\_IV.mov showing the correspondence between the structures viewed from AIA and STEREO A is also provided at the link, which exploits JHelioviewer \citep{mul2017}.}
\end{figure*}

\begin{figure}
\hspace{0cm}\includegraphics[scale=1.15]{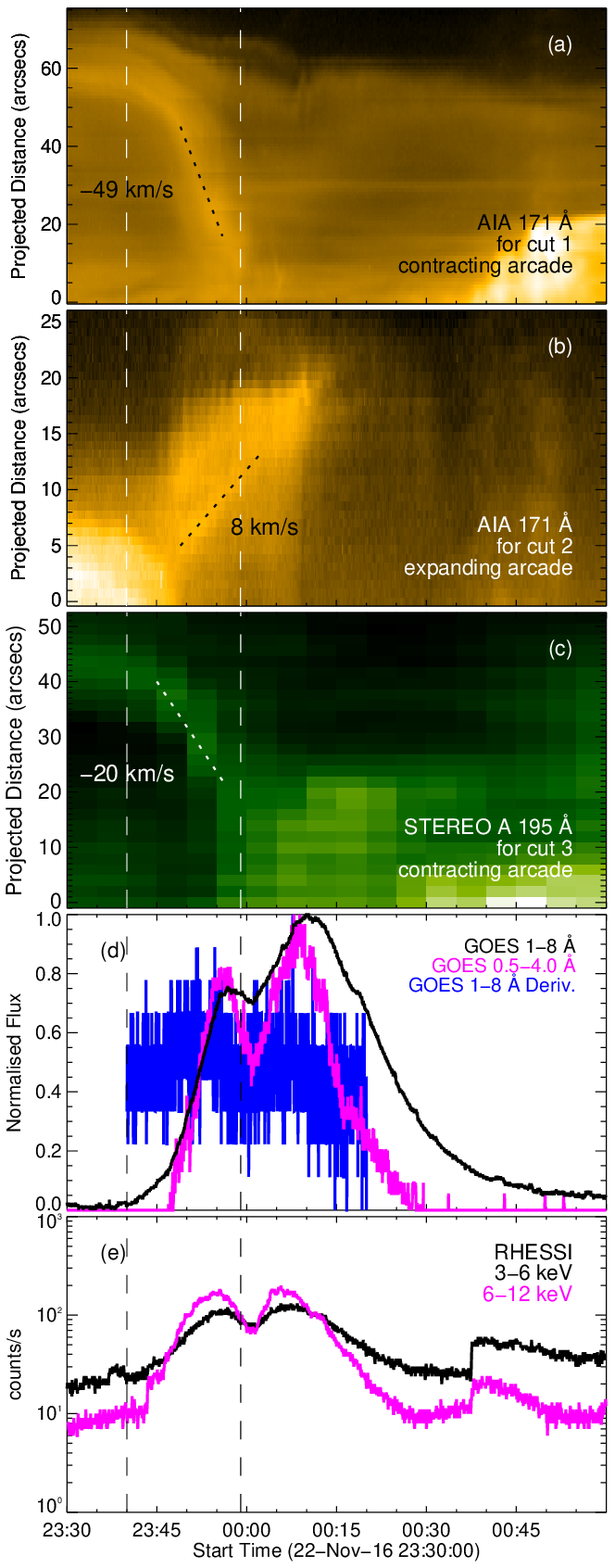}
\caption{\label{ts4}(a)-(c) Timeslices for dynamic features in event IV. The sampling time of STEREO A 195 {\AA} in (C) starts from the beginning of each timeslice, with an exposure duration $\sim8$ s. (d)-(e) GOES and RHESSI light curves, respectively. The two vertical dashed lines across the figure shows the time interval of the arcade contraction.}
\end{figure}

In Event IV, AIA observes two contracting arcade systems with an edge-on geometry (Figure~\ref{cut4}(a) and accompanied animation). Unlike the situation in \cite{zuc2017} where the two peripheral arcades first diverge from each other and then contract, these arcades here directly converge towards each other and contract at the same time (Figure~\ref{cut4}(b)). As they do so, it seems that two flare regions from two sides approach to the convergence location, which may imply that magnetic energy is released gradually towards the central core region. From STEREO A, we also detect the arcade contraction, with a face-on geometry (Figure~\ref{cut4}(e)). The final disappearance of the contracting arcades is also found here without notable oscillation. Similar to Event III, there are no violent arcade or filament eruptions observed by the two instruments, but only a minor arcade expansion in AIA (Figure~\ref{cut4}(b)). From the animation, it appears that this small expansion might be associated with a very weak invisible flux rope erupting outward, or it could also be field line opening due to magnetic reconnection.   

Different from Events I and II, the arcade contraction speed in this event is much larger than the expansion speed (Figure~\ref{ts4}(a) and (b)). More similarities are found between Events III and IV. The contraction distance is much larger than the expansion distance,  and it also happens during the entire impulsive phase (Figure~\ref{ts4}(d) and (e)).

\section{Discussion}\label{discuss}
\subsection{Observational Characteristics}
Unlike previous observations of contracting loops on the solar disk which are plagued by projection effects, we believe that the main contributing factor for the motion of the loops observed from the perspective of SDO/AIA for the four events presented here is real contraction of loops seen approximately edge-on, and argue for it as follows. (1) It seems unlikely that they could actually be tall and narrow loops seen face-on, otherwise the pointed cusp would drag the loop to contract under magnetic tension force even before the event happens, which is not the case in observations. (2) And due to the edge-on property, we can easily exclude the possibility that the shrinking is due to significant loop inclining perpendicular to its plane, though minor changes in inclination can be observed (especially in Events I and II). (3) As large-scale peripheral loops usually have a dipole geometry they would not be expected to bend or distort in their own plane so we can exclude apparent shrinkage due to this (even though in some cases they might happen under the impact of nearby erupting structures, they would restore to their original positions after the eruption completes, which is not observed here; and it is rare to see dramatic loop inclining in its plane when viewed on the solar disk with a face-on geometry; even in Events III and IV, there are no violent eruptions). (4) The last option left to explain the apparent contraction seems to be a real and significant contraction of the loops.

Table~\ref{differ} summarises the relevant information about the four selected events on the large scale. We concentrate on their eruptiveness, dynamic timing, distance and speed, which can separately reflect the onset, duration, total amount and rate of associated energy change. Both Events I and II exhibit violent filament (or arcade) eruptions in close proximity to the contracting arcades (Figures~\ref{cut1}(b) and \ref{cut2}(a)), whereas there are only small expansions of arcades (or at most signatures of very weak, invisible flux rope eruptions) during the arcade contractions for Event III and IV (Figures~\ref{cut3}(b) and \ref{cut4}(b)). The arcades in Event I and II mainly contract at the rise stage of the impulsive phase. By contrast, the arcade contractions respond to their entire impulsive phases in Events III and IV. 

In terms of dynamic timing, distance and speed, Events I and II show the typical characteristics of eruptive flares, with eruption processes prominent in the large-scale dynamics, though the vast majority of eruptive flares are not accompanied by observed arcade contractions like those reported here. Events III and IV seem to have the opposite trend as the arcade contraction process dominates over the expansion/eruption on the large scale. This new type of coronal evolution may present a great challenge to eruptive flare models, like the ``CSHKP'' standard model \citep{car1964,stu1966,hir1974,kop1976} or breakout model \citep{ant1999,aul2000}.

\subsection{Underlying Physics}
What is the physics behind these arcade contraction phenomena? And what causes them to show the two different categories above in Table~\ref{differ}? The implosion conjecture proposed by \cite{hud2000} provides a possible explanation. In his original paper, it was realised that both eruptions and flares as two main approaches to release magnetic energy stored in the corona could cause implosions. As eruptions and flares may involve different evolutionary time scales and large-scale dynamics, naturally we would expect to detect two kinds of implosion processes separately associated with them, characterised by different properties. This analysis raises a likely  interpretation of the two kinds of arcade contraction behaviours observed, i.e., eruption-driven implosions and flare-driven implosions.   

The distinctions between these events in Table~\ref{differ} seem to match this expectation. Violent filament (or arcade) eruptions are seen in Events I and II, dynamically related with the arcade contractions, which may indicate them as eruption-driven implosions. On the contrary, with no such noticeable large-scale eruptions and only flares detected, Events III and IV may represent flare-driven implosions. Supporting evidence comes from the time range during which the contraction happens. In Events III and IV, the arcades contract during the entire impulsive phase, which is expected from the flare-driven scenario, because the flares continually release coronal magnetic energy and reduce the corresponding pressure. However, in Events I and II the major contractions only occur before the peak (or during the rise stage) of the impulsive phase, even though the flares still continue to liberate significant energy in the rest of the impulsive phase. This thus reflects a different responsible source. This could be the associated filament (or arcade) eruptions, as the escape time from the innermost core regions could be shorter than the flare duration. Since in a few well-observed events \cite[][ and Events I and II here]{sun2012,sim2013,wan2016} we notice that the inner loops, closer to the core region, stop contracting almost at the peak of the  impulsive phase, we suggest that it is around this time that the filament escapes from the innermost core region. In the spirit of this argument, the much slower contraction after the major contraction of Event II (Figure~\ref{ts2}(c)) might be interpreted as caused by the ongoing flare just underneath the contracting arcade (see Figure~\ref{cut2} and accompanying animation). The dominance in distance and speed of the eruptions in Events I and II is in accordance with the expectation of the arcade contractions being merely an auxiliary in the global dynamics, whereas the contractions play a more prominent role on the large scale than the expansions/eruptions in Events III and IV, supporting a different triggering source, which could be the flares. Especially, the coincidence of the abrupt acceleration of the contraction and the spike in GOES 1-8 {\AA} flux at $\sim$ 01:20 UT in Event III (Figure~\ref{ts3}) implies a close connection between these two phenomena.

\begin{table*}
\caption{Focused Large-scale Properties of the Four Selected Events \label{differ}}
\tablewidth{0pt}
%\tabletypesize{\scriptsize}
\centering
\begin{tabular}{c|p{3.cm}|p{3.cm}|p{3.cm}|p{3.cm}}
\hline\hline
& \multicolumn{1}{>{\centering}p{3.cm}|}{SOL2011-09-14T16:26} & \multicolumn{1}{>{\centering}p{3.cm}|}{SOL2014-02-17T23:15} & \multicolumn{1}{>{\centering}p{3.cm}|}{SOL2016-04-08T01:56} & \multicolumn{1}{>{\centering}p{3.cm}}{SOL2016-11-22T23:45} \\
 & \multicolumn{1}{>{\centering}p{3.cm}|}{Event I} & \multicolumn{1}{>{\centering}p{3.cm}|}{Event II} & \multicolumn{1}{>{\centering}p{3.cm}|}{Event III} & \multicolumn{1}{>{\centering}p{3.cm}}{Event IV} \\
 \hline
\multirow{2}{*}{Eruptiveness} & \multicolumn{2}{>{\centering}p{6cm}|}{possess visible, significant filament (or arcade) eruptions} & \multicolumn{2}{>{\centering}p{6cm}}{only have small and weak arcade expansions; no obvious filament (or arcade) eruptions} \\
\hline
\multirow{2}{*}{Timing} & \multicolumn{2}{>{\centering}p{6cm}|}{mainly contract during the rise stage of the impulsive phase} & \multicolumn{2}{>{\centering}p{6cm}}{contract during the entire impulsive phase} \\
\hline
\multirow{4}{*}{Distance} & \multicolumn{2}{>{\centering}p{6cm}|}{arcade contraction distance (Event I: $\sim10$ arcsec; Event II: $\sim20$ arcsec) is much smaller than filament (or arcade) eruption distance (Event I: $>70$ arcsec; Event II: $\sim200$ arcsec)} & \multicolumn{2}{>{\centering}p{6cm}}{arcade contraction distance (Event III: $\sim40$ arcsec; Event IV: $\sim45$ arcsec) is much larger than arcade expansion distance (Event III: $\sim15$ arcsec; Event IV: $\sim15$ arcsec)} \\
\hline
\multirow{5}{*}{Speed} & \multicolumn{2}{>{\centering}p{6cm}|}{arcade contraction speed (Event I: $\sim45~ \rm km~s^{-1}$; Event II: $\sim100~ \rm km~s^{-1}$) is much smaller than filament (or arcade) eruption speed (Event I: $\sim221~ \rm km~s^{-1}$; Event II: $\sim246~ \rm km~s^{-1}$)} & \multicolumn{2}{>{\centering}p{6cm}}{arcade contraction speed (Event III: $\sim5~ \rm km~s^{-1}$; Event IV: $\sim49~ \rm km~s^{-1}$) is comparable to, or much larger than arcade expansion speed (Event III: $\sim4~ \rm km~s^{-1}$; Event IV: $\sim8~ \rm km~s^{-1}$)} \\
\hline
Possible Origin & \multicolumn{2}{>{\centering}p{6cm}|}{eruption-driven implosions} & \multicolumn{2}{>{\centering}p{6cm}}{flare-driven implosions} \\
\hline
\end{tabular}
% \tablenotetext{a}{Adjust}
\tablecomments{For Events III and IV, the expanding structures could incline toward or away from SDO, resulting in underestimations of their travelling distances and speeds, but from the accompanied animations and geometry, it seems that they do not incline too much. If we assume the inclination angle to be a characteristic value $\sim45$\degree, the conclusions here still hold, not to mention that the contracting structures could not be in the sky plane as well.}
\end{table*}

\subsection{Models}\label{model}
\begin{figure*}
\centering
\includegraphics[width=0.8\textwidth]{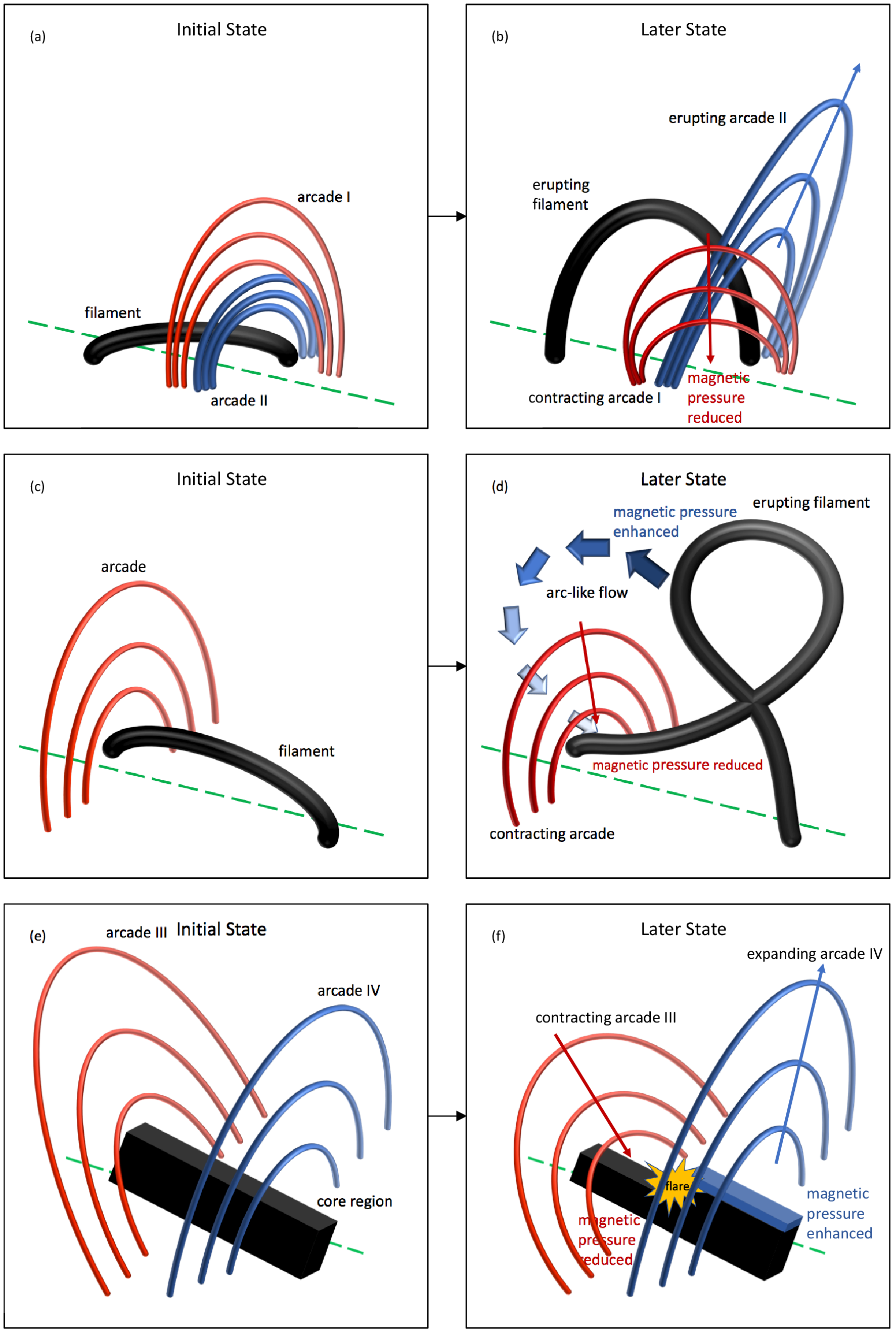}
\caption{\label{cartoon}Cartoons show our understanding of the implosion events. (a)-(b) for Event I. (c)-(d) for Event II. (e)-(f) for Events III and IV. The thin arrows in each image indicate the directions of the implosion and expansion motions of the arcades. And the green dashed line represents the polarity inversion line.}
\end{figure*}

Figure~\ref{cartoon} illustrates our understanding of these four events exploiting the implosion conjecture. Figure~\ref{cartoon}(a)-(b) and Figure~\ref{cartoon}(c)-(d) describe the field evolution of Events I and II, respectively. As argued above, Events I and II are of eruption-type, thus possessing similar essential dynamic characteristics, i.e., when the underlying filament erupts outward, the peripheral overlying arcade contracts. This scenario is also used to interpret the event in \cite{wan2016}. The basic idea is that filament (or arcade) field redistribution, and/or conversion of its energy to kinetic and gravitational energy, can locally reduce magnetic energy and pressure in its original position, resulting in forces in the periphery being unbalanced and the associated loops contracting. Another interesting explanation by \cite{zuc2017} and \cite{dud2017} is that the eruption and contraction in this MHD situation are an analog of a fast flow creating vortices in its surroundings in hydrodynamics. However, due to the preferable perspectives here, we see that, in Event I (Figure~\ref{cut1} and accompanying animation) arcade I just adjacent to arcade II contracts directly when arcade II erupts, without the significant initial expansion and inclination phases that are expected in the vortex-flow scenario \citep{dud2017}. And in Event II the arcade only shows an arc-like flow rather than a complete vortex trajectory in the hydrodynamic situation, which is also illustrated in Figure~\ref{cartoon}(d). In theory, the viscous term in the invoked momentum equation \citep{zuc2015,aul2005} of the simulation performed by \cite{zuc2017} and \cite{dud2017} is much smaller than the Lorentz force in a low $\beta$ coronal MHD environment. Thus, the viscosity, which is responsible for vortex generation in the hydrodynamic case, would not be able to create the large-scale organised rapid contraction behaviours, though it might produce small-scale vortices around the erupting structure. The large-scale dynamics is controlled by the dominant Lorentz force. \cite{zuc2017} argued that it is the enhanced magnetic tension, one component of the Lorentz force, caused by compressional Alfv\'{e}n waves originating from the erupting field, that generates the contraction flow, but according to this argument, the contracting loops are expected to restore to their original locations after the filament (or arcade) erupts completely because of the nature of waves, which does not agree with the reported observations in which the loops remain at lower altitudes. Similarly, if the contracting motion was only caused by enhanced magnetic pressure (the other component of the Lorentz force) above the loops due to the erupting structure, we would also expect their restoration when the eruption terminates, not conforming to the observations either.

The final idea then resorts to reduced magnetic pressure underneath, which is just the core idea of the implosion conjecture. In fact, the arc-like flow in Figure~\ref{cartoon}(d) can be easily explained in this framework. As the filament erupts outward, the magnetic pressure is enhanced at higher altitude and reduced at lower altitude, which would naturally induce an arc-like flow of peripheral unopened arcade field around the central erupting structure because of pressure difference compared to the previous equilibrium state. Depending on the detailed topology and eruption process, the arc-like flow may not be so obvious in some cases, like Event I here; and the loops located at lower altitudes where they are not severely impacted by the high-pressure erupting structure could also contract directly, e.g., the event in \cite{sim2013}. The perturbation in the pressure should propagate outward with a limited speed, as observed by \cite{sim2013} in a face-on geometry. This could be the fast-mode speed ($\sim$ Alf\'{v}en speed $v_A$ if plasma $\beta\ll1$ as in the corona).

Particularly, there is strong observational evidence that Events III and IV do not show violent eruptions and vortex-like or even arc-like flows. The arcade in Event III contracts directly, and the two arcades of Event IV even converge towards each other and simultaneously contract downward. The contractions are significantly different from peripheral vortices created by a central fast flow in hydrodynamics, and thus cannot be explained by the analogy. Instead, the implosion conjecture \citep{hud2000} is able to account for these two events, in terms of flare-driven implosions, without the need for eruptions. This has already been supported by the distinct properties of Events III and IV in Table~\ref{differ}, as argued above. Because of a mix of difficulties from limb location, structure overlapping in an edge-on geometry, and low contrast, the 3D field topologies of Events III and IV are not readily reconstructed. However, we propose a general model for them to interpret the major contractions and minor expansions observed, based on the implosion conjecture. Figure~\ref{cartoon}(e)-(f) illustrate the basic idea. The ``black box'' underlying the two arcade systems represents the core region where a flare occurs. During the flare the total magnetic energy and pressure are reduced within the entire ``black box''. However, there could exist a situation where the field energy underneath arcade III decreases and that underneath arcade IV increases, but the increase under arcade IV is smaller than the decrease under arcade III. Then we would expect to see that the contraction of arcade III is larger in extent and faster in speed than the expansion of arcade IV, which would then be in agreement with the properties of Events III and IV in Table~\ref{differ}. However, the detailed field reconnection process, corresponding topology change and energy transport and dissipation in the ``black box'' are unclear . The magnetic energy enhancement underneath arcade IV might be due to more closed field formed or field opening there through reconnection between the two domains under the two arcade systems. Such a model of flare-driven implosions is attractive and can reproduce the observations in a general way, but another possibility, which cannot be completely excluded, is that a small and invisible flux tube may continuously transport from under arcade III toward arcade IV, in the spirit of eruption-type implosions but a very weak one.

\subsection{Unsuccessful Implosion}
It is worth noting that well-observed implosions, either face-on or edge-on remain rather rare, whereas the implosion conjecture implies that they should be present in all solar energy-releasing events, including eruptions and flares. This is probably because of unfavourable viewing, complexity of the active region field and reconnection processes involved \citep{liu2010}, or relatively small expected movements in readily observed peripheral loops when relatively small fraction of active region energy is released in the core region in a flare. However, in this context we would like to revisit one of the original assumptions for the implosion conjecture in \cite{hud2000}, i.e., that gravity takes no significant role in the coronal dynamics. This might not always be the case, especially when a filament is involved, and this could lead to unsuccessful implosions. Take the illustrations Figure 9(c)-(d) for example in a general way (rather than considering the specific Event II). Suppose, as a thought experiment, that before the eruption in Figure 9(c), the filament is mass loaded, with the downward gravitational force contributing a non-negligible amount to the force balance against the upward Lorentz force. Now imagine what would happen if simultaneous with the MHD instability much of the material along the filament drained down to the photosphere. As the local plasma density and thus gravitational pull are reduced, the filament field would inflate, simultaneously pushing the overlying arcade outward, which is the opposite of an implosion. Similarly, during the eruption in Figure 9(d), such a process would occur if mass along the filament field could drain down \cite[see relevant studies, e.g.,][pointing out that substantial filament material drains down that may influence the dynamics]{fon2002,bi2014,fan2017,jen2018} and also spread into a larger volume. Moreover, as the filament field becomes more vertical, the draining could increase, further inflating the surrounding field. Thus the overlying arcade would expand if the magnetic energy change associated with the filament is not considered. However, in fact, the filament field becomes ``weaker'' locally, distributing into a larger volume and transferring its energy into plasma kinetic and gravitational energy. As argued by \cite{hud2000} and \cite{rus2015}, to achieve a new equilibrium, the overlying arcade would implode toward the magnetic-pressure-reduced filament. At the end, in this scenario we would have two competing mechanisms controlling the dynamics: gravity reduction making the field expand and magnetic pressure reduction making the field implode. In some cases, the magnetic pressure reduction is dominant so we see implosions, like Events I and II here, while the gravity reduction may overtake in other situations, which might be one of the reasons for rarity of well-observed implosions.

\section{Conclusions}\label{conclusion}
With the four selected events having the up-to-now most clearly observed continuously contracting loops in an edge-on geometry from the viewpoint of SDO/AIA, supplemented by observations from STEREO, for the first time we demonstrate the existence of real contractions of loops in the global coronal dynamics unambiguously. The implosion conjecture proposed by \cite{hud2000} in the interpretation of these events is found to be effective, in comparison with alternative theories for which disagreements currently exist between observations and simulations or other predictions. Meanwhile, the discussion also leads us to find two implosion categories that can be associated either with solar eruptions or with flares, and the models put forward according to the conjecture can reasonably explain their distinct observational characteristics. However, it is also pointed out that in some cases the implosion scenario may not be valid as one of the original assumptions about the role of gravitation in the dynamics may fail.

\acknowledgments
 The authors thank the referee for very helpful comments and suggestions, which significantly improve the quality of the manuscript, and Hugh Hudson for discussion. L.\,F. and P.\,J.\,A.\,S. acknowledge support from STFC Consolidated Grant ST/L000741/1 and ST/P000533/1. P.\,J.\,A.\,S. acknowledges support from the University of Glasgow's Lord Kelvin Adam Smith Leadership Fellowship. The authors are grateful to NASA/SDO, AIA, HMI, STEREO/EUVI, and GOES science teams for data access.
\facilities{SDO(AIA and HMI), STEREO(EUVI), GOES}.

\bibliography{implosion2}

% \begin{thebibliography}{}
% 
% 
% \end{thebibliography}

\end{document}